\documentclass{PoS}

\title{Peculiar nature of hard X-ray eclipse in SS433 from INTEGRAL observations}

\ShortTitle{Hard eclipse of SS433}

\author{\speaker{A. Cherepashchuk}$^{a}$, K. Postnov$^{a}$, 
E. Antokhina$^{a}$, S. Molkov$^{b}$\\
        \llap{$^{a}$}Sternberg Astronomical Institute, 119992 Moscow, Russia\\
	\llap{$^{b}$}CESR, Toulouse, France\\
        E-mail: \email{cherepashchuk@gmail.com}, 
\email{kpostnov@gmail.com}, \email{elant@sai.msu.ru}, \email{Sergey.Molkov@cesr.fr}}

\abstract{The analysis of hard X-ray INTEGRAL observations (2003-2008) of 
superaccreting galactic microquasar SS433 at precessional
phases of the source with the maximum disk opening 
angle is carried out. It is found that the shape and width of the primary 
X-ray eclipse is strongly variable suggesting additional absorption
of X-ray photons in dense stellar wind and gas outflows from the
optical A7I-component. The joint modeling of X-ray eclipse
and precessional X-ray variability of SS433 revealed by INTEGRAL 
using a 
geometrical model suggests the binary mass ratio 
$q=m_x/m_v\simeq 0.3$. 
This value of the binary 
mass ratio allows us to easily explain peculiarities of the optical
variability of SS433, in particular, the substantial 
precessional variability of the minimum brightness at the middle
of the primary optical eclipse. 
For the mass funciton of
the optical star $f_v=0.268 M_\odot$ as derived from Hillwig \& Gies data \cite{hillwig08}
the obtained value of $q$ 
yields the masses of the components $m_x\simeq 5 M_\odot$, 
$m_v\simeq 15 M_\odot$, confirming the black hole nature of
the relativistic object in SS433. 
The independence of the observed hard X-ray spectrum on the accretion disk
precession phase suggests that hard X-ray emission ($kT=20-100$~keV)
is formed in an extended, hot, quasi-isothermal corona, probably 
heated by interaction of relativistic jet with 
inhomogeneous wind outflow from 
the precessing supercritical accretion disk.
The Monte-Carlo simulations of broadband X-ray spectrum of SS433
at the maximum disk opening precessional phases 
allowed us to determine physical parameters of the 
hot corona (temperature $T_{cor}=20$~keV, Thomson optical depth  
$\tau=0.2$), as well as to estimate the mass outflow  rate in jets $\dot M_j=3\times 10^{19}$~g/s yielding the kinetic power of the jets $\sim 10^{39}$~erg/s.
}

\FullConference{7th INTEGRAL Workshop\\
		 September 8-11 2008\\
		 Copenhagen, Denmark}

\begin{document}

\section{Introduction}
\label{sec:intro}

SS433 is a massive close binary system at andvanced evolutionary stage
\cite{Margon84, Cher88, Fab04}. This unique galactic X-ray
binary with 
precessing relativistic jets ($v=0.26c$, where $c$ is the speed of light)
exhibits several variabilities, including the precessional one (with the
period 
$P_{prec}\simeq 162$~d), the eclipsing one (with the binary orbital period  
$P_{orb}=13.08$~d), and the nutational one (with the nutation period
$P_{nut}\simeq 6.28$~d) (e.g. \cite{Goransk98}). SS433 is recognized 
as a galactic microquasar with precessing supercritical accretion disk
around a relativistic compact object, and has been extensively investigated 
in the optical, radio and X-ray ranges (for a comprehensive
review and references see \cite{Fab04}). The optical spectroscopy of the system
\cite{hillwig08} revealed the presence of absorption lines in the
spectrum of the optical component identified as a $\sim$~A7I supergiant
star. Observed orbital Doppler shifts of the absorption lines of the optical
component allowed Hillwig \& Gies to determine the mass ratio of the relativistic
($m_x$) and the optical ($m_v$) components in SS433 $q=m_x/m_v\simeq
0.35$, implying the masses $m_x=4.3\pm0.8 M_\odot$ and $m_v=12.3\pm 3.3
M_\odot$. Similar masses were obtained from our analysis of the optical
light curves of SS433 \cite{AntCher87}. 

Until now, the analysis of X-ray eclipses in SS433 in the 1-10 keV range
has yielded controversial results suggesting a small mass ratio $q\simeq
0.15$ (e.g. \cite{Kawai89}). The main reason for this
is a very broad X-ray eclipse observed in this X-ray band. 

Our studies of SS433 in hard X-rays using the INTEGRAL observations
suggested the presence of a hot rarefied corona above the supercritical
accretion disk in this source \cite{Cher03, Cher05, Cher06}. The
peculiar variability in the shape and width of the primary eclipse in hard
X-rays was discovered. This implies that the primary eclipse is not purely 
geometrical and that the binary mass ratio as derived from the eclipse
duration may be unrealistic. 

The results of our analysis of the oprbital and precessional variability in
SS433 observed by INTEGRAL \cite{Cher06} can be summarized as
follows. The hard X-ray flux (25-50 keV) from the source clearly exhibits
the precessional variability with the precessional period $P_{prec}=162.4$~d
from $\sim 3$~mCrab at the cross-over phase to $\sim 18-20$~mCrab at the
maximum disk opening phase (the $T_3$ moment, where the moving emission
lines in the SS433 spectrum are at maximum separation). The orbital
eclipses are observed with the orbital period $P_{orb}=13.08$~d and are very
significant. Eclipses observed close to the $T_3$ phase (at the precessional phase $\psi\simeq 0.1$) are the deepest ones. The hard X-ray flux (18-60 keV) at the center of a primary eclipse is detectable at a level of $\sim 3$~mCrab, so the ratio of the maximum uneclipsed flux ($\sim 20$~mCrab) to the minimum value at the mid-eclipse is about 6-7. The width of hard X-ray eclipse is found to be bigger than that in soft X-rays. The egress out of the hard X-ray eclipse is observed to be strongly variable, most probably due to absorption of the X-ray flux by accretion flows and asymmetric, structured wind emanating from the supercritical accretion disk. Similar distortions of the eclipse egress was first observed by Ginga (18.4-27.6 KeV) \cite{Kawai89}. The hard X-ray spectrum (20-200 keV) does not noticeably change with the precession phase. All these facts suggest that the hard X-ray flux of SS433 is mostly generated in the hot extended corona formed in the central parts of the accretion disk. A detailed interpretation of the borad-band (3-90 keV) X-ray continuum of SS433 in terms of the multicomponent model including the accretion disk, jet and corona has been carried out by \cite{Krivosheev08}. 

In the present paper we analyse hard X-ray eclipses of SS433 near the T3 moment
as observed by INTEGRAL and interprete them in terms of our multicomponent 
geometrical model with account of the peculiar shape and strong variability of the primary eclipse. 

\section{Precessional variability}

Dedicated INTEGRAL observations of 
SS433 were carried out in AO-1 for 500 ks, in AO-3 for 500 ks, in AO-4 in May 2007 for 466 ks, 
and in AO-5 for $\sim 900$ ks. 
The AO-1 observations (67-69 INTEGRAL orbits) were performed 
near the T3 moment (near the precessional phase 0 with maximum disk opening). 
Out of the eclipse, the source is reliably detected 
by IBIS/ISGRI up to 100 keV, 
with the X-ray spectrum fitted by featureless power-law $dN/dt/dA/dE\sim 
E^{-2.8}$~ph/s/cm$^2$/keV \cite{Cher03, Cher05, Cher06}, see Fig. \ref{Xspectra}.
Considering a low significance of points above $\sim 100$ keV, 
the exponential cut-off in the spectrum is still not excluded.
The spectral continuum 2-100 keV  
can be explained by thermal emission from cooling expanding plasma jet with temperature at the jet base $T_0\sim 22$ keV, and a broad hot comptonized region with $T_c\sim 20$ keV and optical
depth $\tau\sim 0.2$  
surrounding the thin X-ray jet \cite{Krivosheev08}.  

In AO-2, SS433 fell within the FOV of IBIS when observing the Sagittarius
Arm Tangent region, but no X-ray eclipses occurred during this program. 

In AO-3, SS433 was observed around different precessional phases (INTEGRAL orbits 366-369). 
One X-ray eclipse was partially measured again at the precessional phase close to zero, with an indication of much narrower eclipse or 
a sudden mid-eclipse (at $\psi_{orb}\sim 1.03$) flux increase. 
Based on these data, the model for the source eclipse has been 
constructed in \cite{Cher05, Cher06}; however, 
these data were not strongly constrained by precessional variability and 
allowed a broad range of the binary mass ratio $q\sim 0.1-0.5$. 

In A0-4, the source was observed in May 2007; unfortunately, due to high variability, 
the source flux was very low, so we cannot include these data in the analysis of
X-ray eclipses (see set II in Fig. \ref{f_alleclipses}). Note the strong increase of
the eclipse width. 

In AO-5, two consecutive eclipses of the source near the zero precessional phase 
were observed in October 2007. Adding all of these observations allowed the spectroscopy at different precessional phases  
(Fig. \ref{Xspectra}). However, the statistics is still unsufficient to 
make orbital phase-resolved spectroscopy. 

The combined
AO1 -- AO5 data (including those publically available) 
enabled us to measure, for the first time, the precessional 
hard X-ray variability (Fig. \ref{f1}), which 
turned out to be quite significant and stable over several 
precessional periods. The maximum to minimum flux ratio of the precessional variability 
was found to be around 4, which is higher than in softer X-ray bands, and evidences for 
the hard X-ray emission originating closer to the basement of the visible part of the jets.  
The eclipsing and precessional variabilities, combined with 
spectroscopic data simultaneously taken with INTEGRAL observations by the 6-m telescope SAO RAS, were taken into account when attempting to model the light curve of the source by \cite{Cher05}. 
However, the results 
proved to be inconclusive. First of all, the spectral resolution of the optical observations (around 3000) was insufficient to definitely measure the radial velocity curve of the optical star, and 
the modeling of X-ray eclipse and precessional variability allowed fairly broad range of parameters, including the mass ratios $q\sim 0.1-0.3$. An independent spectral analysis of archival  RXTE observations of SS433 \cite{Fil06} also suggests that it is hard to evaluate the mass ratio
based on X-ray observations only.

\section{Hard X-ray spectra}

\begin{figure}
\includegraphics[width=0.5\textwidth]{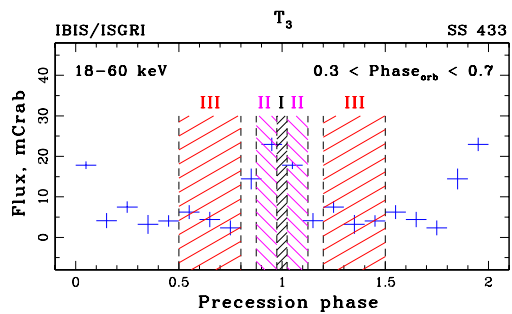}
\caption{Precessional phase intervals chosen for spectral analysis of \SS433.
\label{f1}}
\end{figure}

\begin{figure}
\includegraphics[angle=270,width=0.3\textwidth]{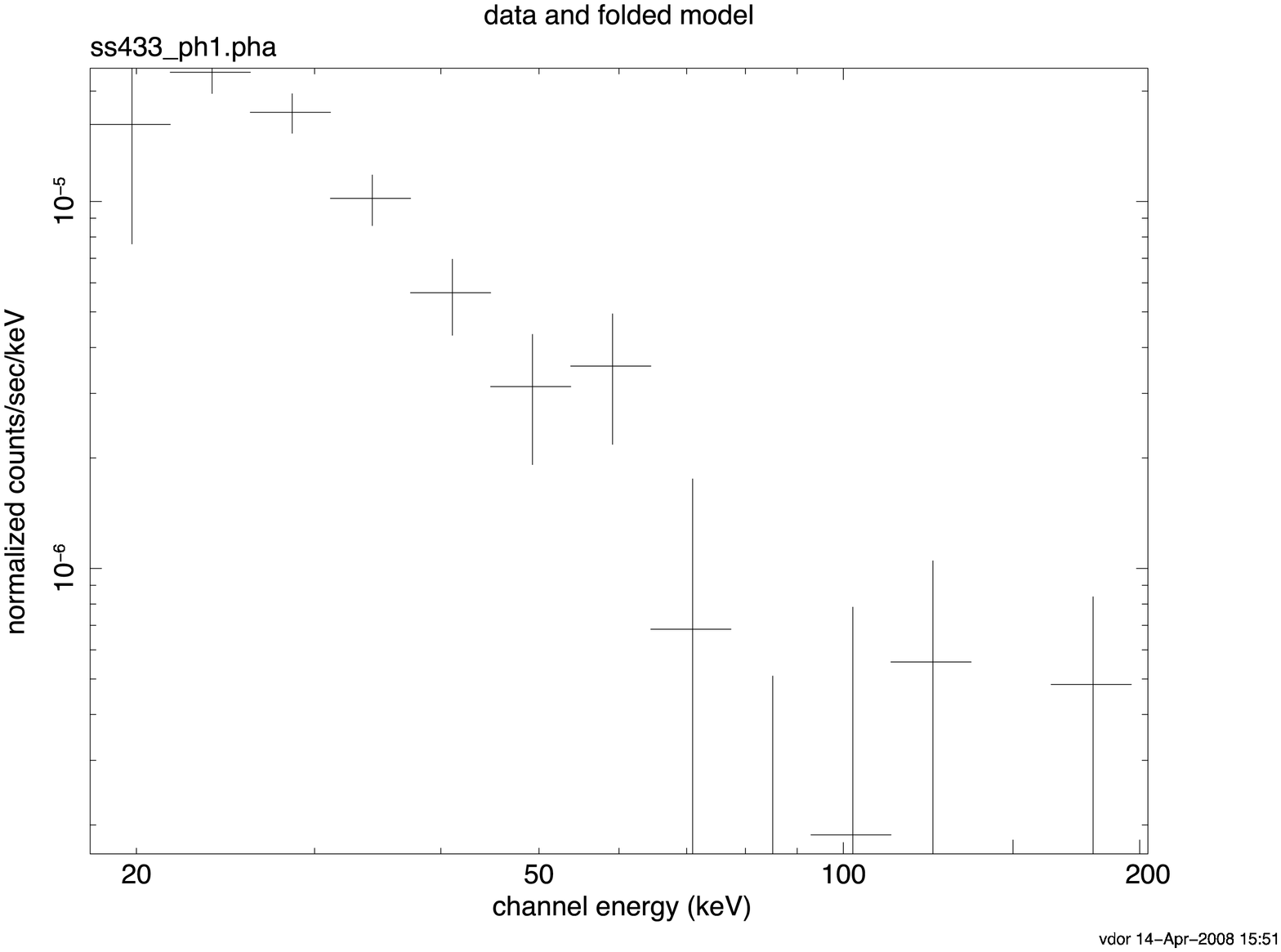} 
\hfill
\includegraphics[angle=270,width=0.3\textwidth]{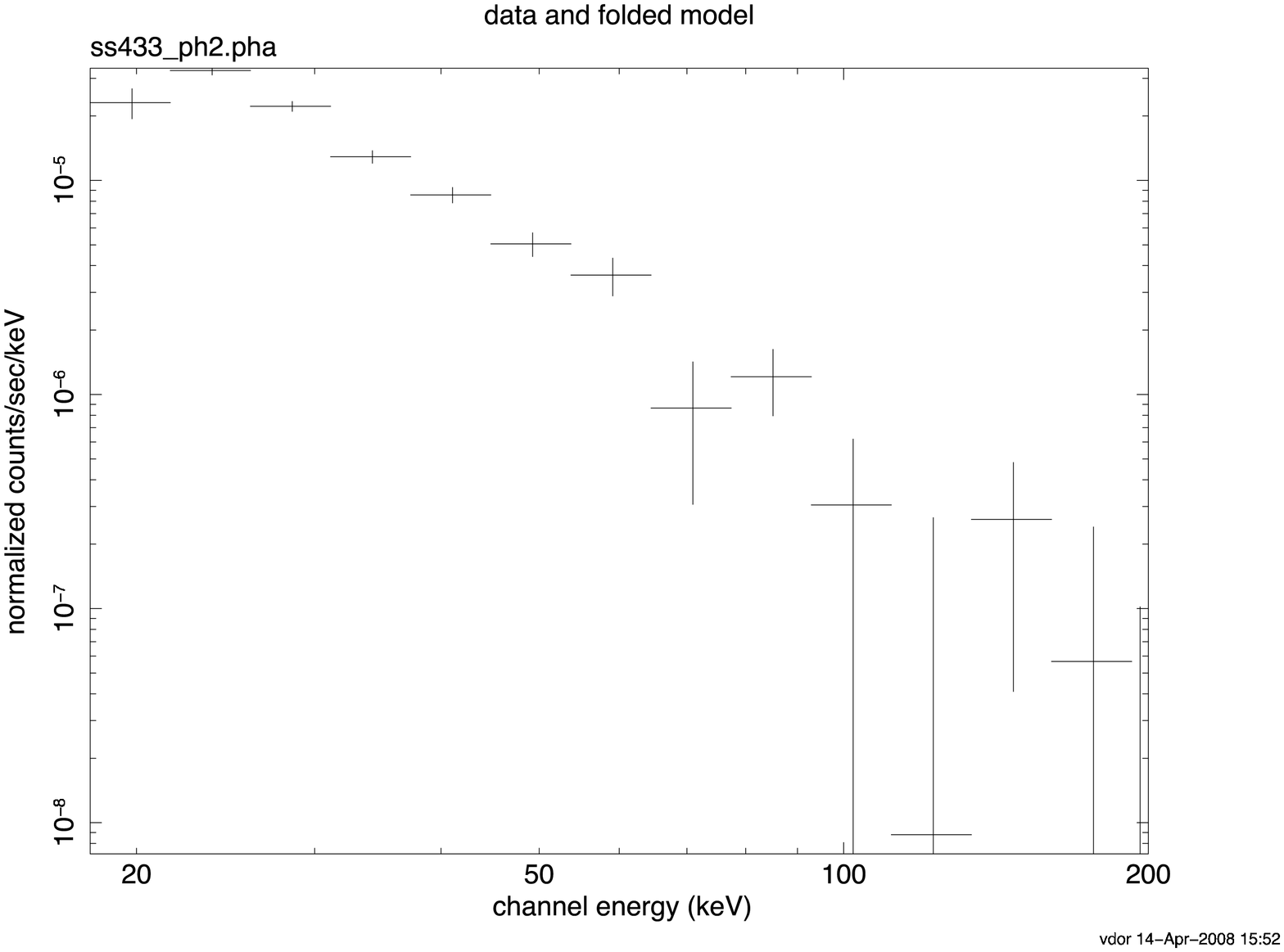} 
\hfill
\includegraphics[angle=270,width=0.3\textwidth]{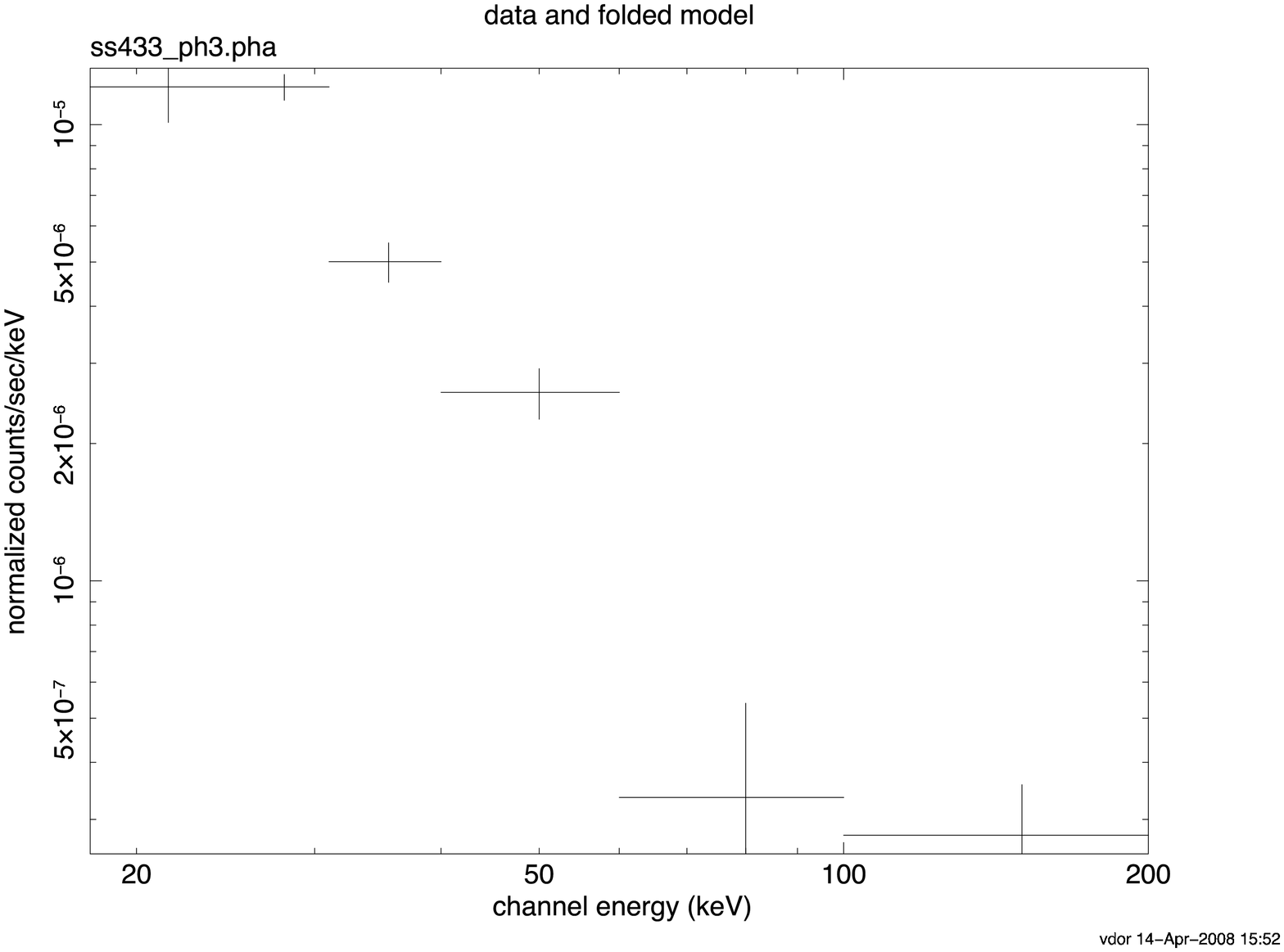} 
\caption{IBIS/ISGRI $10-200$~keV SS433 spectra at different precessional phases (I, II, III in Fig. 1). All spectra can be fitted by a power-law model with the photon spectral index $-2.8$.}
% \parbox[7]{0.47\textwidth}{\caption{Eclipse fit for the mass ratio $q=0.3$.}\label{eclipse}}
% \hfill
% % \parbox[7]{0.47\textwidth}{\caption{Precessional (25-50 keV) light curve of SS433 from INTEGRAL observations.}  
\label{Xspectra}
\end{figure}

Our previous studies \cite{Cher03,Cher05,Cher06} revealed that SS433 is 
a superaccreting galactic microquasar with hot corona above the accretion 
disk. SS433 is only one known massive X-ray binary in which such a corona is directly observed. 
To find spectral signatures of the hot corona, 
INTEGRAL observations (2003-2007) were separated into 
three segments in the precessional phase (Fig. \ref{f1}) (limited by photon statistics). 
The resulted spectra are shown in Fig. \ref{Xspectra}, with no clear difference in the power-law spectral shape.
This confirms the presence of 
a fairly broad region emitting in hard X-rays with size 
compared to that of the accretion disk ($10^{11}-10^{12}$ cm), 
since (excluding orbital eclipses) 
the observed flux varies due to precession of the disk.

Guided by the constant shape of the hard X-ray spectrum, we summed up all precessional 
phases (to increase statistics). 
The obtained X-ray spectrum of SS433 in the 3-100 keV range 
can be fitted by two-component model (thermal X-ray emission from 
the jet and thermal comptonization spectrum from corona) 
elaborated in \cite{Krivosheev08}.
% (Fig. \ref{Xmodel}).  
The model gives the best fit
% is shown 
for $T_{cor}=20$ keV with Thomson optical depth $\tau_T=0.2$
and mass outflow rate in the jet $\dot M_j=3\times 10^{19}$~g/s.

% \begin{figure}[ht]
% \includegraphics{compdom.eps,width=0.47\textwidth} 
% \hfill
% \includegraphics[width=0.47\textwidth]{obssim.eps} 
% \parbox[7]{0.47\textwidth}{\caption{Computational domain for Monte-Carlo calculation of hard X-ray spectrum from corona and jet.}}
% \hfill
% \parbox[7]{0.47\textwidth}
% {\caption{3-100 keV spectrum of SS433 (including normalized JEM-X data) fitted by two-component jet+corona model. See \cite{Krivosheev08} for details.}  
% \label{Xmodel}}
% \end{figure}

The best-fit corona parameters
obtained from spectral fits suggest the electron density around $5\times 10^{12}$ cm$^{-3}$. 
Such a density is typical in the wind 
outflowing with velocity $v\sim 3000$ km/s from a supercritical 
accretion disk with $\dot M\sim 10^{-4}$
M$_\odot$/yr at distances $\sim 10^{12}$ cm from 
the center, where a Compton-thick photosphere is formed \cite{Fab04, Revnivtsev04}.

Note that despite several observations of primary eclipses by INTEGRAL, the
observed hard X-ray flux $\sim 5-20$ mCrab from the source is still insufficient 
for phase-resolved spectral analysis of an individual eclipse. Additional
observations of SS433 by INTEGRAL are planned to allow phase-resolved 
spectroscopy of the primary eclipse. 

\section{Primary eclipse and its peculiarity}
\subsection{INTEGRAL eclipses}

The INTEGRAL data were processed with both publically available
ISDC software (OSA-7 version) and the original software
elaborated by the IKI INTEGRAL team (for the IBIS/ISGRI telescope,
see \cite{Mol04} for more detail).

\begin{figure}
\includegraphics[width=0.47\textwidth]{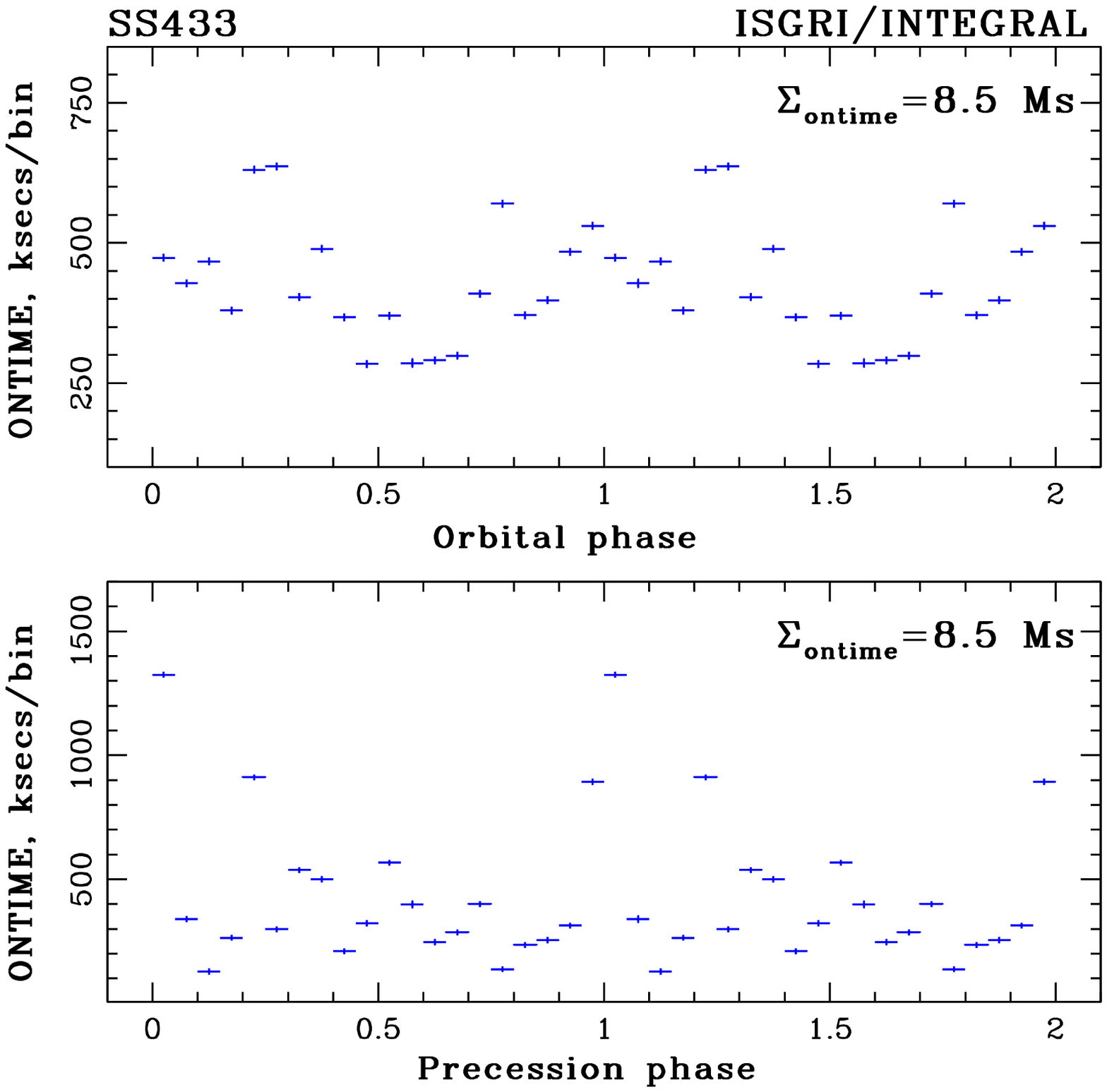}
\hfill
\includegraphics[width=0.47\textwidth]{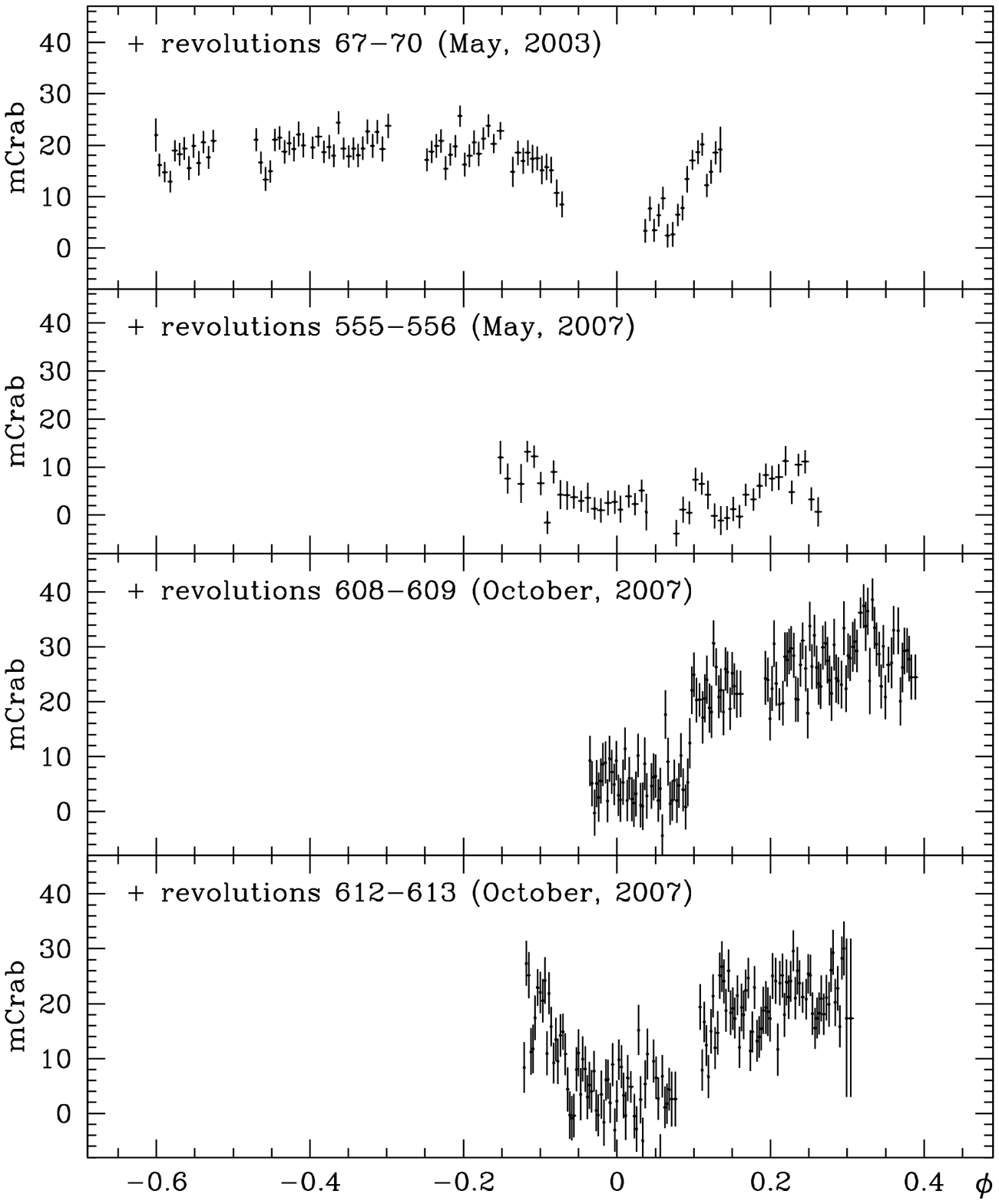}
\parbox[t]{0.47\textwidth}
{\caption{Distribution of the observing time from the orbital (upper) and the precession (bottom)
phases.\label{distr_tot_exp}}}
\hfill
\parbox[t]{0.47\textwidth}
{\caption{The INTEGRAL primary eclipses of SS433 (IBIS/ISGRI data, 18-60 keV) I, II, III and IV sets (see the text).\label{f_alleclipses}}}
\end{figure}

In our analysis we have used data from our INTEGRAL observing programm of
SS433 plus publically availabe data of all observations where the
source was in the FOV of the IBIS/ISGRI telescope (<13 deg). The
exposure of selected data total an approximately $8.5$ Ms. To perfome
precessional phase-resolved analysis we ascribed to each SCW (Science Window
or SCW, natural piece of INTEGRAL's data -- pointing observation with
exposure $\sim2-5$ ks) appropriate orbital and precessional phases.
The phases are calculated using the ephemeris provided by \cite{Fab04},
and modifyed for INTEGRAL time data format. According to these ephemeris
the relative orbital phase is: 
$$
\Psi_{orb} = (T_{IJD}-1520.880)/13.08211
$$
and the relative precessional phase is :
$$
\Psi_{prec}= (T_{IJD}-8037.03)/162.375
$$
In Fig. \ref{distr_tot_exp} (left panels) we shown the exposure distribution of
used observations with the phases. For clarity we presented two cycles.
On the orbital phase panel ``1'' corresponds to the orbital minimum,
and on the precessional phase panel ``1'' is the $T3$ moment. While the
exposure time for the orbital phases is distributed more or less
homogeneously that in the precessional phase plane has excess near
$T3$ moment. The reason is that the main part of our SS433 INTEGRAL
observing programm were focused namely on observations of primary eclipses. 

In this paper we focused on the precessional phase-resolved analysis.
To exclude contamination of the orbital modulation on the results, in 
the present analysis 
we have used only observations with orbital phases from the range
$0.3<\Psi_{orb}<0.7$. After this filtering we still have $\sim2.8$~Ms
of data distributed across all precession phases.
% as presented in Fig. \ref{distr_tot_exp} (to the right).  

% \begin{figure}
% \includegraphics[width=0.47\textwidth]{expdistr_0.3-0.7.ps}
% % \caption{Distribution of the observing time on the precession phases. The data
% with $\Psi_{orb}<0.3$ and $\Psi_{orb}>0.7$ were excluded.}
% \label{prec_exp_distr}
% \end{figure}

The dedicated INTEGRAL observations of primary eclipses of SS433 in 2003 and 2007 were 
obtained near the T3 
precession phase when the accretion disk is maximally open. The data include:

 I. INTEGRAL orbits 67-70 (May 2003), precession phase $\psi_{pr}=0.001-0.060$;

 II. INTEGRAL orbits 555-556 (May 2007), precession phase $\psi_{pr}=0.980-0.014$;

 III. INTEGRAL orbits 608-609 (October 2007), precession phase $\psi_{pr}=0.956-0.990$;

IV. INTEGRAL orbits 612-613 (October 2007), precession phase $\psi_{pr}=0.030-0.064$. 

\noindent The phases are calculated using the ephemeris \cite{Fab04},
according to which the orbital minimum is 
$$
JD_{MinI} (hel) = 2450023.62 + 13.08211*E\,,
$$
the T3 moment is 
$$
JD_{T3}= 2443507.47 + 162.375*E1\,.
$$
The observed eclipses are shown in Fig \ref{f_alleclipses}. In our analysis
we have used only sets I, III, and IV and excluded set II (the one showing 
the most suppressed egress out of the primary eclipse).  

\subsection{Variable shape of hard X-ray eclipse}

\begin{figure}
\includegraphics[width=0.6\textwidth]{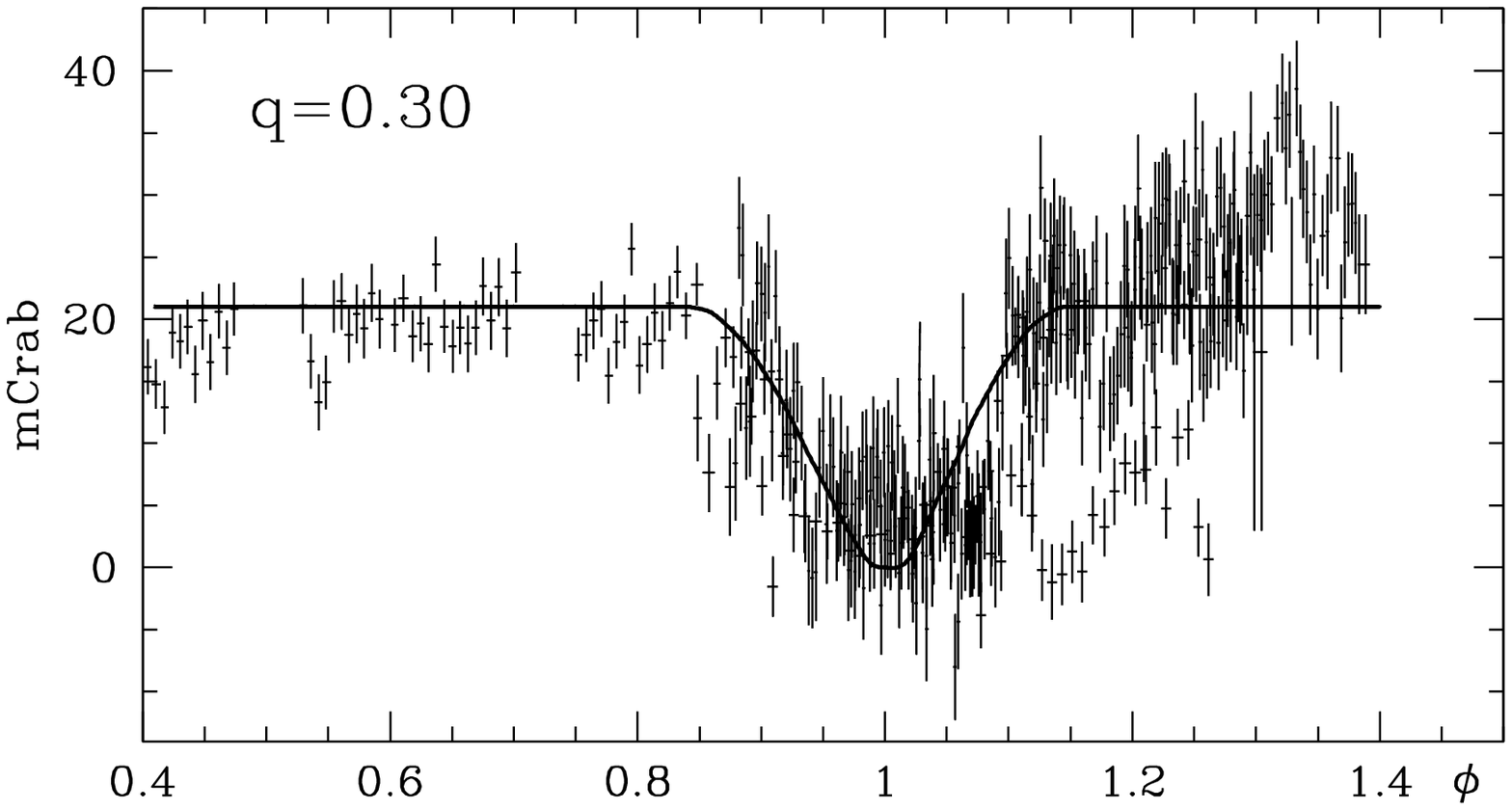}
\hfill
% \parbox[]{0.5\textwidth}{
%\includegraphics[width=0.37\textwidth]{scheme_and_model.eps}
\includegraphics[width=0.37\textwidth]{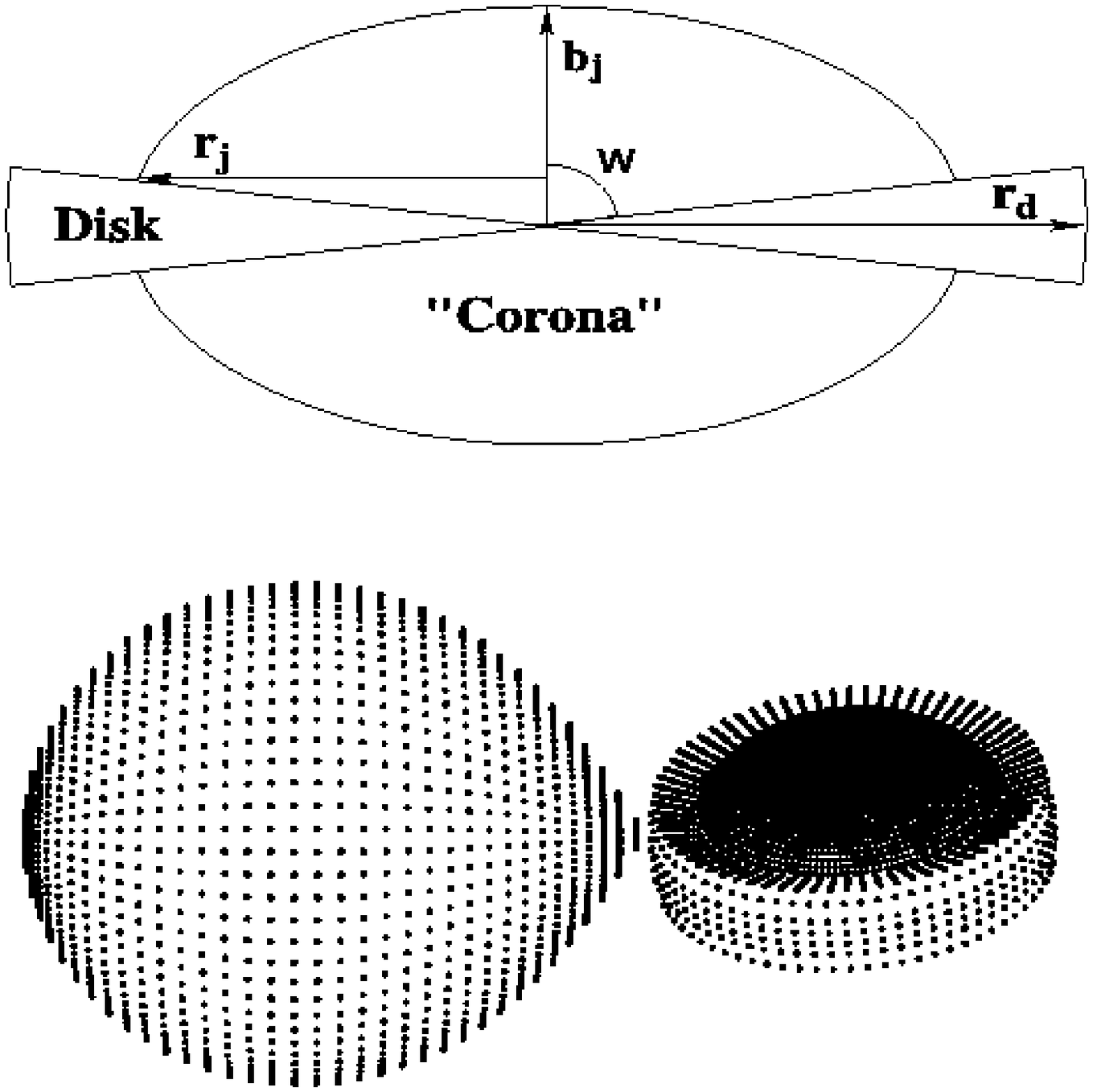}
% \vfill
% \includegraphics[width=0.5\textwidth]{fig2E_model_scheme.eps}
% }
%\centering \includegraphics{cartoon.eps}
\parbox[t]{0.6\textwidth}{\caption
{Combined INTEGRAL primary eclipses of SS433. The solid
curve shows the model light curve for $q=0.3$ (see the text).}\label{eclipse}}
\hfill
\parbox[t]{0.37\textwidth}{\caption{Geometrical model of the accretion disk and its "corona" (top) and the view of the SS433 geometrical model for $q=0.3$ at the maximum
disk opening angle (bottom).}\label{f_geommodel}}
\end{figure}

SS433 is known to be highly variable object at all wavelengths \cite{Cher88,Fab04,Revnivtsev04,Revnivtsev06}. This is not surprizing for a supercritically
accreting massive binary, considering various instabilities that should be immanently present in the
mass outflow from the optical star and in the accretion disk. 
It might appear that the primary eclipse of the hard X-ray emission 
coming from the regions close to the 
central black hole by the optical star should be stable. However,
it is not the case for hard X-ray eclipse in SS433 (Fig. \ref{f_alleclipses}). As seen in this figure, the second eclipse observed in May 2007 has an unusually broad form: a shallow ingress to and broad egress from the eclipse, which may be  due to the low level of uneclipsed X-ray flux from the source $\sim 10$~mCrab. Note that this particular eclpse is 
very similar to the one observed by \textit{Ginga} \cite{Kawai89}. However, 
already at the next T3 moment in October 2007 (sets III and IV in Fig. \ref{f_alleclipses}) the uneclipsed X-ray flux was at a level of 20 mCrab and the egress from eclipse restored its more familiar form (like in set I). 
This effect is particulary clearly visible on the combined X-ray eclipse light curve shown in Fig. \ref{eclipse}. The reason for such a strong variability of the hard X-ray egress may be
related to powerful inhomogeneous gaseous streams feeding the disk and the wind from 
the supercritical accretion disk. Indeed, the slow dense disk wind in SS433 
should form a Compton-thick 
photosphere around the disk and funnels around relativisatic jets \cite{Fab04}. The height of the 
funnel $\sim 10^{12}$, comparable to the accretion disk size, 
can be inferred from the observed $\sim 80$-s lag of the optical emission with respect to the X-rays from the jet base found in the simultaneous optical/X-ray observations of the source
\cite{Revnivtsev04}. So it appears quite possible that the variable wind from the precessing supercritical accretion disk 
result in changes of the funnel structure, so even at the maximum opening disk
phase (T3) the shape of hard X-ray eclipse may appreciably vary depending 
on the current accretion rate through the disk. In this connection it is worth
noticing the variable broad absorption-like feature that is seen immediately after
the egress from the eclipse (at the orbital phases $\psi_{orb}\sim 0.15-0.2$, 
see Fig. \ref{f_alleclipses}). It would be very interesting to confirm if this
feature is due to the true absorption of X-ray emission or just reflects the strong variability of the proper X-ray flux. 

Therefore we conclude that the hard X-ray eclipse in SS433 is likely to be formed by 
both geometrical screening of the broad X-ray emitting region by the opaque
star and by complex gaseous stream and/or inhomogeneous wind from the star and the supercritical accretion disk. This implies that  
to analyse the hard X-ray primary eclipse in SS433 by geometrical model 
we should use the most stable part of the observed eclipses, i.e. the upper
envelope of the observed eclipse ingress. Moreover, 
in view of high variability, the use of only one X-ray eclipse for analysis
can lead to erroneous determination of the binary system parameters.  

\subsection{Geometrical model of SS433}

To analyse hard X-ray eclipses of SS433 we used a geometrical model 
developed earlier
to the interpretation of the {\it Ginga} data \cite{Ant92}
and the {\it INTEGRAL} light curve \cite{Cher05}.
We consider a close binary system consisting of an (opaque) "normal"  star
limited by the Roche equipotential surface and a relativistic object
surrounded by an optically and geometrically thick "accretion disk".
Relativistic jets are directed perpendicularly to the disk plane.
The "accretion disk" includes the disk itself and an extended photosphere
formed by the outflowing wind. The orbit is circular, the axial rotation
of the normal star is assumed to be synchronized with the orbital revolution. 

% \begin{figure}
% \centering\includegraphics[width=0.5\textwidth]{fig2E_model_scheme.eps}
%\centering \includegraphics{cartoon.eps}
% \caption{Geometrical model of the accretion disk and its "corona".}
% \label{f_geommodel}
% \end{figure}

The disk and "jets" are precessing in space and change the orientation
relative to the normal (donor) star. 
The disk is inclined with respect to the orbital plane by the angle $\theta$.
A cone-like funnel is located inside the disk and is characterized by the
half-opening angle $\omega$, thus the opaque disk body (see Fig. \ref{f_geommodel}) is
described by the radius $r_d$ and the angle $\omega$.
The central object is surrounded by a transparent 
homogeneously emitting spheroid with a visible radius $r_j$ and height $b_j$
which could be interpreted as a "corona" or a "thick jet" (without any
relativistic motion). Here $r_j$, $b_j$ and $r_d$ are dimensionless values
expressed in units of the binary separation $a$.
The radius of the normal star is determined by the relative Roche lobe size,
i.e. by the mass ratio $q=m_x/m_v$ ($m_x$ is the mass of the
relativistic object).

Only the "corona" is assumed to emit in the hard X-ray band, while the star
and disk eclipse it in the course of the 
orbital and precessional motion. During
precession the inclination of the disk with respect to the observer changes,
causing different visibility conditions for the "corona".
The "corona" is partially screened by the cone disk edge, which can
explain qualitatively the change of the uneclipsed X-ray flux with
precession phase. 
Observations of
precessional variability can thus be used to obtain a "vertical" scan of the
emitting structure, resctricting the parameters $b_j$  and $\omega$.
The orbital (eclipse) variability observations scan the emitting
structure "horizontally", restricting possible values of $r_d$, $\omega$, $q$ and
$r_j$. The joint analysis of the precessional and eclipse variability enables us
to reconstruct the spatial structure of the region in the accretion disk
center where the hard X-rays are produced and to estimate the binary mass ratio $q$.

The position of the components of the system relative to the 
observer is determined by the binary orbit inclination angle $i=78.8^o$, 
the disk inclination angle to the orbital plane $\theta=20.3^o$, and 
the precessional phase $\psi_{pr}$. Phase $\psi_{pr}=0$ corresponds to the maximum 
disk opening of SS433 (T3 moment, maximum separation between the moving emission
lines) and $\psi_{pr}=0.34, 0.66$  when the disk is seen edge-on 
(at the moving emission line crossover moments T1 and
T2, respectively). 

\subsection{Light curves analysis}

Free parameters of our model for the orbital and precessional variability of SS433 include: the binary mass ratio $q=m_x/m_v$ determining the relative 
size of the normal star, 
the disk parameters $r_d, \omega$ determining the form of the disk, 
the thick "jet" or "corona" parameters $r_j, b_j$ determining the form of the hot corona. Note that the X-ray emitting region is qualitatively different for 
a thin narrow "jet" with $r_j\ll b_j\sim a$ and a thin short "jet" with $r_j\ll  b_j\ll a$. If $r_j>b_j$, it is more appropriate to term it as a corona (or
a thick "jet"). For each value of $q$ from the range $0.05-1.0$, we found other parameters best-fitting simultaneously the orbital 
and precession light curves. The precession variability amplitude was
assumed to be $A_{pr}\simeq 5-7$. The $\chi^2$ criterion was used to evaluate
the goodness of fits. 

To examine as fully as possible all system parameters, we have used 
different variants of the observational data modeling. These included:

1) the analysis of light curves consisting of individual points;

2) the analysis of average light curves;

3) the analysis of individual observational sets (I, III, and IV);

4) the analysis of the combined data from sets I, III and IV;

5) the analysis of the entire eclipse, including the ingress and egress phases;

6) the analysis of only ingress to the eclipse (orbital phases $\psi_{orb}=0.8-1.0$).

The last analysis appears to be the most reliable, since all observations
of the primary eclipse of SS433 evidence that the form of 
egress out of eclipse is strongly variable and thus cannot be described by purely geometrical model of eclipses.

Our numerous calculations led to the following conclusions. 

1) The results obtained using individual and average light curves are
identical (i.e. the using of the average light curve does not add errors
to the obtained parameters). This might seem to be obvious, but we 
directly checked it. 

2) The results obtained from the analysis of individual data sets I, III, IV are the same as ones obtained from the analysis of the combined data. 

Taking into account the above conclusions, below we shall show the model
light curve for the combined data sets I, III and IV.

3) When searching for the best-fit model parameters at different $q$, 
we found that solutions obtained for the entire primary eclipse and for 
the eclipse ingress only are not very different. However, taking into account highly
variable character of the eclipse egress noted above, below we shall discuss solutions and model parameters obtained from the analysis of the eclipse ingress part only. 

4) In our previous analysis of the primary orbital eclipse observed by INTEGRAL \cite{Cher05} we found that for all models the best-fit values 
(corresponding to the minimum $\chi^2$) are obtained for a maximum possible radius of the disk $r_d$ and a maximum possible value of the angle $\omega$. This conclusion is confirmed here using new data. We also confirm that parameters $b_j$ and $\omega$ are correlated. 

\begin{figure}[ht]
\includegraphics[width=0.47\textwidth]{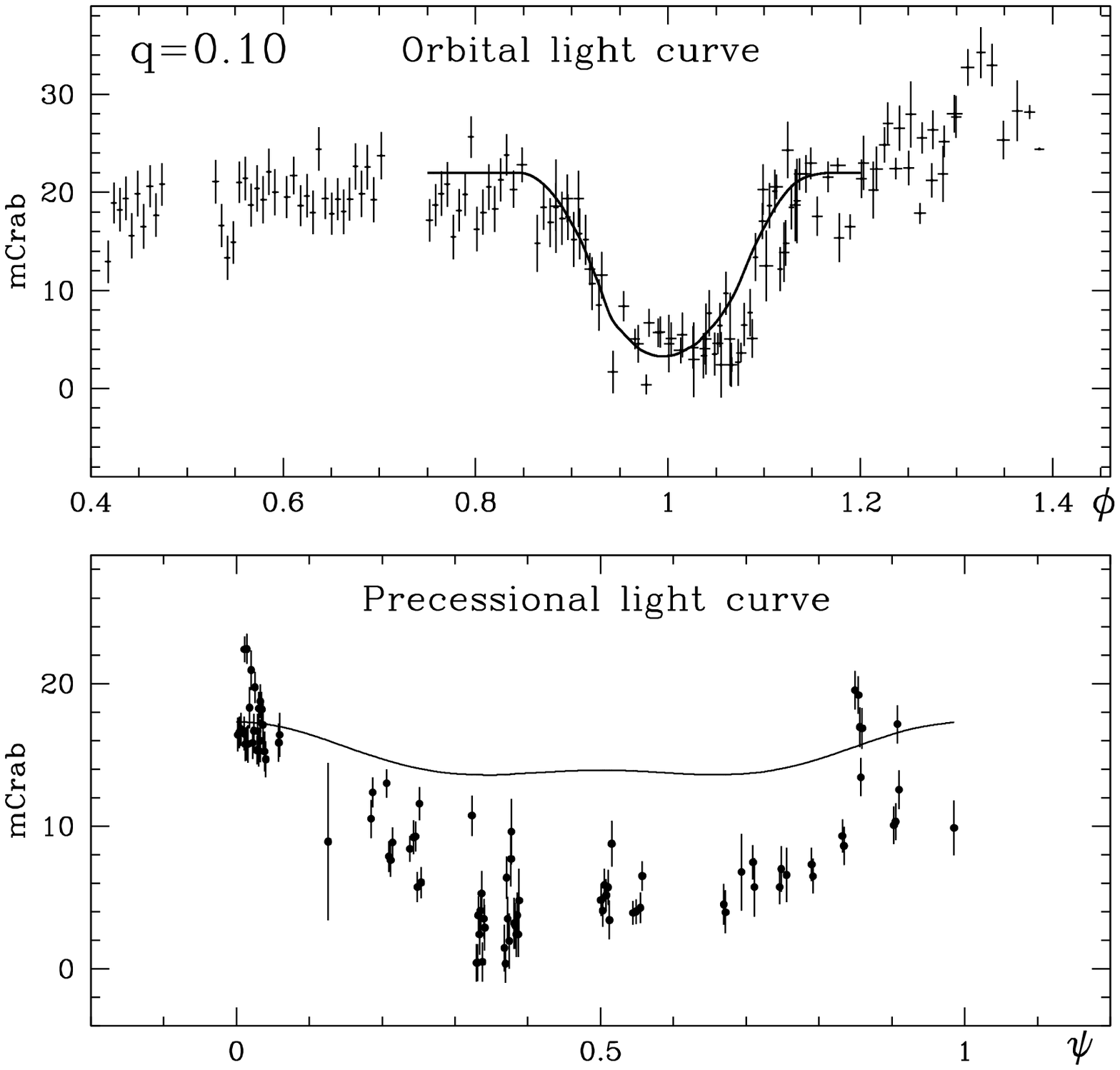}
\hfill
\includegraphics[width=0.47\textwidth]{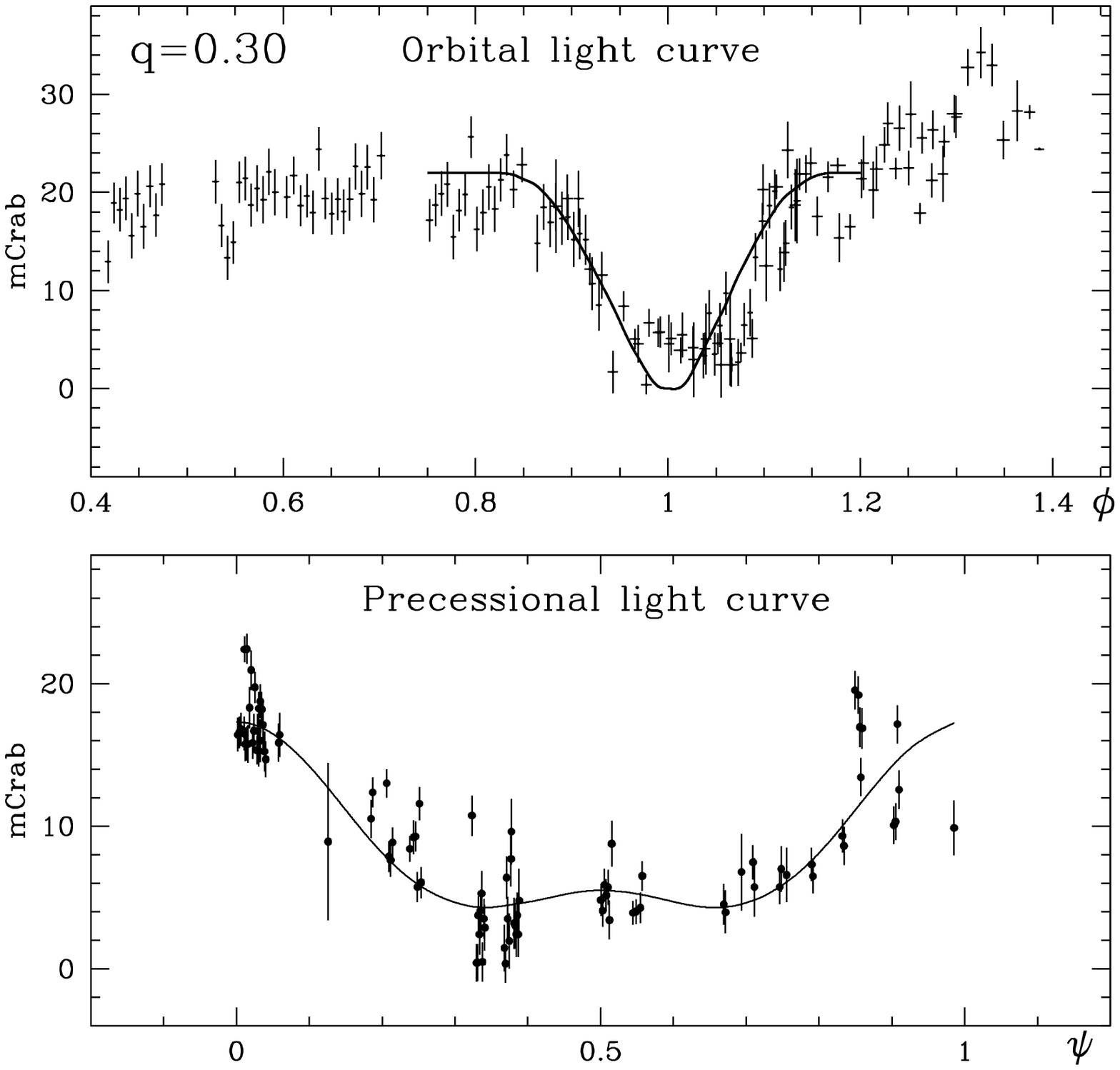} 
\parbox[t]{0.47\textwidth}{\caption{Orbital (top) and precessional (bottom) light curves of SS433 from INTEGRAL observations for model with $q=0.1$. The long  
jet model is shown (see the text). The model can not simultaneously fit
the eclipse and precessional light curves.}  
\label{f_longjet}}
\hfill
\parbox[t]{0.47\textwidth}{\caption{X-ray eclipse and precessional light curve of
SS433 for the mass ratio $q=0.3$ 
Note highly variable eclipse egress. The solid curve shows the model eclipsing and
precessional light curve (see the text).
\label{q03}}}
\end{figure}

5) The minimum deviation of the best-fit model eclipsing light curve from observed points is reached for the small mass ratio $q\sim 0.1$ and a long X-ray emitting "jet" ($b_j>0.5$) with the base radius varying in the wide range ($r_j=0.05-0.25$). This means that the "jet" can be long but either thin or thick. However, the main objection to model of the long "jet" comes from the impossibility to describe
the observed precessional variability. 
This is illustrated in Fig. \ref{f_longjet}. 
We conclude that at small mass ratio $q\le 0.2$ our
model does not simultaneously fit both orbital and precessional variability of SS433 observed by INTEGRAL. The precessional light curve could be fitted at the small mass ratio by a short X-ray emitting "jet", but then the flux at the center of the primary eclipse would be zero, again  
% (Fig. \ref{f_longjet}, right panel), 
in contradiction 
to observations -- in all cases, INTEGRAL detects a non-zero X-ray flux of $\sim 3$~mCrab at the middle of hard X-ray eclipse (Fig. \ref{f_alleclipses}).

% \begin{figure}[ht]
% \includegraphics[width=0.5\textwidth]{fig3E_q01_long_jet.ps} 
% \hfill
% \includegraphics[width=0.5\textwidth] {fig4E_q01_short_jet.ps}
% \parbox[7]{0.47\textwidth}{\caption{Best-fit to the orbital light curve by long X-ray jet and
% small binary mass ratio $q=0.1$.}}
% \hfill
% % \caption{Orbital and precessional light curve of SS433 from INTEGRAL observations for model with $q=0.1$. The long  
% and short jet model are shown in the left and right panel, respectively.
% For both models it is impossible to simultaneously fit
% the eclipse and precessional light curves.}  
% \label{f_longjet}
% \end{figure}

% \begin{figure}
% \centering \includegraphics{fig_model_scheme.eps}
%\centering \includegraphics{cartoon.eps}
% \caption{Best-fit to the orbital light curve by short X-ray jet at 
% small binary mass ratio $q=0.1$ (right panel). The precessional variability 
% is shown on the left panel. Note the zero flux expected at the center of the primary eclipse.}
% \label{f_shortjet}
% \end{figure}

6). For $q=0.3-0.5$ our model provides simultaneously good fits to the orbital and precessional variability of SS433, although $\chi^2$ deviations are formally worse than for small values of $q$. The deviations increase with $q$, so the minimal $\chi^2$ are obtained for $q=0.3$. The model parameters 
correspond to an extended corona comparable to the accretion disk in width ($r_j\sim r_d$) and with small vertical size ($b_j\sim 0.15-0.20$). This
geometry of the hard X-ray emitting region is caused by the wide primary eclipse minimum and large amplitude of the precessional variability. So the precessional light curve is shaped by the outer parts of the disk eclipsing 
the high-temperature corona. 
Fig. \ref{q03} 
% and \ref{q05} 
shows the orbital and precessional 
light curves for the best-fit binary mass ratio $q=0.3$. 
% (left panel) and $q=0.5$ (right panel). 
Figs. \ref{f_longjet} and \ref{q03} clearly demonstrate 
the change in the primary eclipse minimum width for different binary mass ratios. 

\subsection{Masses of the components}

From the analysis of the primary eclipse at the precessional phase $\Psi=0.1$ (the upper envelope) with account for precessional variability we have obtained the mass ratio $q=m_x/m_v\simeq 0.3$. 
Using the mass function of the optical component of SS433 derived from 
spectroscopic observations by \cite{hillwig08}, 
% Hillwig and Gies (2008) 
\begin{equation}
f_v(m)=\frac{m_x\sin^3i}{(1+1/q)^2}\simeq 0.268 M_\odot\,,
\end{equation}
we get masses of the relativistic object and the optical star:
\begin{equation}
m_x\simeq 5 M_\odot, \qquad m_v\simeq 15 M_\odot\,.
\end{equation}
The high mass value of the relativistic object in SS433 strongly suggests its being a black hole. 

\section{Discussion}

Our analysis confirms the nature of SS433 as a superaccreting microquasar with black hole. The obtained high value of the binary mass ratio in SS433 $q=0.3$ allows us to easily explain a substantial amplitude ( $\sim 0^m.5$) of the optical variability at the minimum of eclipse with precessional phase. The orbital inclination of SS433 is known from independent  analysis of moving emission lines ($i\simeq 78^\circ.8$), so at small $q\simeq 0.1\div 0.2$ the small relative radius of the Roche lobe of the relativistic object should have caused the total eclipse of the bright precessing accretion disc around the relativistic object by the optical A7I-star, implying a constant optical flux at the center of the eclipse at different precession phases. This contradicts to observations (see e.g. \cite{Goransk98}). At $q=0.3$ the size of the Roche lobe of the relativistic object is relatively large to cause partial primary optical eclipse of the accretion disc by the A7I-star at all precession phases, implying significant precessional variability of the minimum brightness at the middle of the primary eclipse. 

A small value of the binary mass ratio was found by \cite{Kawai89}
% Kawai et al. (1989) 
%and  \cite{Kotani96}
% Kotani et al. (1996)  
from the analysis of soft ($kT\simeq 1\div 10$~keV) X-ray eclipses of SS433. As follows from the results of modeling of broad-band X-ray spectrum of SS433 \cite{Krivosheev08}, in this energy range thermal emission of relativistic jets with temperature decreasing along the jet dominates. The observed broad width of soft X-ray eclipse was interpreted by the authors \cite{Kawai89},
% (Kawai et al. 1989, Kotani et al. 1996), 
by applying the purely geometrical model of the eclipse of jets by the optical star with sharp limb,  found the small value of the binary mass ratio $q=0.15$. The observed independence of the soft X-ray eclipse width on energy was the main argument justifying this model and low mass ratio. However, this argument holds only if the temperature  along jets does not change with distance from the relativistic object. In fact it decreases along jets (unless some additional heating mechanism is assumed), so at energies $\sim 10$~keV mainly central parts of the jets are eclipsed. Consequently, the duration of purely geometrical eclipses must be longer at 10 keV than at $\sim 1$~keV where cooler periphery of the jets is eclipsed \cite{Fil06}.  The independence of the X-ray eclipse duration on energy of X-ray photons in the 1-10 keV range can be due to the compensation of the effect of decreasing temperature along jets and the increase of soft X-ray absorption in the extended atmosphere of the optical star. So the effective radius of the eclipsing star increases at smaller energy of X-ray photons which are mostly emitted at smaller temperature along the jets. This explains the constant width of X-ray eclipse in the 1-10 keV energy range. Turning this argument around, from the independent width of the X-ray eclipse in this energy range we can conclude that the eclipsing A7I-star in SS433 has not a sharp limb, so that X-ray eclipses in SS433 are not purely geometrical and are suffered from extinction in the stellar wind from the optical star. This lends additional support to our interpretation of X-ray eclipses in SS433 which makes use of only the upper envelope of variable eclipsing hard X-ray light curve. 

\section{Conclusion}

The discovery of highly variable hard X-ray eclipse at precessional phases corresponding to the
maximum disk opening angle for the observer and
significant precessional variability are 
the main findings of our INTEGRAL observations of SS433. 
It is shown that the duration of the primary X-ray eclipse in 
different epochs changes by $\sim 2$ times. This implies that 
X-ray eclipses in SS433 are not purely geometrical, they are shaped
by absorption in a powerful stellar wind and gas streams from the optical star. So to infer the binary mass ratio $q$ from the 
analysis of X-ray eclispses we have used only the upper envelope 
of the eclipsing light curve. The analysis of this upper envelope 
in combination with the precessional light curve yielded the 
mass ratio estimate $q=0.3$. The relatively high value of the binary 
mass ratio provides an easy explanation of peculiarities of the optical
variability of SS433, in particular, the substantial 
precessional variability of the minimum brightness at the middle
of the primary optical eclipse. 

Using the mass function of the optical star found by \cite{hillwig08} $f_v(m)=0.268 M_\odot$ 
and the value of $q=m_x/m_v\simeq 0.3$ inferred from our analysis, we concluded that 
the masses of binary components of SS433 are $m_x\simeq 5 M_\odot$, 
$m_v\simeq 15 M_\odot$. The high mass of the relativistic 
objet leaves no doubts that it is a black hole. 

The independence of hard X-ray spectrum on the accretion disk
precession phase suggests that hard X-ray emission ($kT=20-100$~keV)
is formed in an extended, hot, quasi-isothermal corona. The heating
of the corona can be due to transformaiton of kinetic energy of relativistic jets to an inhomogeneous wind outflow from 
the precessing supercritical accretion disk \cite{Begelman06}.

The Monte-Carlo simulations of broadband X-ray spectrum of SS433
at the maximum disk opening precessional phases \cite{Krivosheev08}
allowed us to determine the main physical characteristics of the 
hot corona (temperature $T_{cor}\simeq 20$~keV, Thomson optical depth  
$\tau\simeq 0.2$), as well as to estimate the mass outflow  rate in jets $\dot M_j=3\times 10^{19}$~g~s$^{-1}$ yielding the kinetic power of the jets $\sim 10^{39}$~erg~s$^{-1}$.

\acknowledgments
The authors thank M. Revnivtsev for useful
discussions. The work is
partially supported by the RFBR grants 07-02-00961 and 08-02-01220.

\end{document}